\documentclass{aa}
\usepackage{txfonts}
\usepackage{epsfig}
\usepackage{subfigure}
\usepackage{multirow}

\usepackage{times}
\usepackage{natbib}
\usepackage{url}
\bibpunct{(}{)}{;}{a}{}{,}

\begin{document}

\title{The X-ray spectral signatures from the complex circumnuclear regions in the Compton thick AGN NGC~424}

\author{Andrea Marinucci\inst{1}, Stefano Bianchi\inst{1}, Giorgio Matt\inst{1}, Andrew C. Fabian\inst{2}, Kazushi Iwasawa\inst{3}, Giovanni Miniutti\inst{4}, Enrico Piconcelli\inst{5}}

\offprints{Andrea Marinucci\\ \email{marinucci@fis.uniroma3.it}}

\institute{Dipartimento di Fisica, Universit\`a degli Studi Roma Tre, via della Vasca Navale 84, 00146 Roma, Italy
\and Institute of Astronomy, Madingley Road, Cambridge CB3 0HA, UK
\and ICREA Research Professor at Institut de Ci\`encies del Cosmos, Universitat de Barcelona, Mart\'i i Franqu\`es, 1, 08028 Barcelona, Spain
\and Centro de Astrobiolog\'{i}a (CSIC--INTA), Dep. de Astrof\'{i}sica; LAEFF, P.O. Box 78, E-28691, Villanueva de la Ca\~{n}ada, Madrid, Spain
\and Osservatorio Astronomico di Roma (INAF), Via Frascati 33, I-00040 Monte Porzio Catone, Italy
}

\date{Received / Accepted }

\authorrunning{A. Marinucci et al.}
\titlerunning{The X-ray spectral signatures from the circumnuclear regions in the Compton Thick AGN NGC~424}

\abstract 
{} 
{Most of our knowledge of the circumnuclear matter in Seyfert galaxies is based on the X-ray spectra of the brightest Compton-thick Seyfert 2 galaxies. The complete obscuration of the nuclear radiation in these sources allows us to study all the components arising from reprocessing of the primary continuum in the circumnuclear matter in detail, while they are heavily diluted in unobscured sources, often down to invisibility.}
{We present the XMM-\textit{Newton} RGS and EPIC pn spectra of a long ($\simeq100$ ks) observation of one of the soft X-ray brightest Compton-thick Seyfert 2 galaxies, NGC 424. As a first step, we performed a phenomenological analysis of the data to derive the properties of all the spectral components. On the basis of these results, we fitted the spectra with self-consistent photoionisation models, produced with \textsc{cloudy}.}
{The high-energy part of the spectrum is dominated by a pure neutral Compton reflection component and a neutral iron K$\alpha$ line, together with K$\alpha$ emission from neutral Ni, suggesting a significant Ni/Fe overabundance. The soft X-ray RGS spectrum comes mostly from line emission from H-like and He-like C, N, O, and Ne, as well as from the Fe L-shell. The presence of narrow RRC from \ion{O}{viii}, \ion{O}{vii}, and \ion{C}{vi}, the last two with resolved widths corresponding to temperatures around 5-10 eV, is a strong indication of a gas in photoionisation equilibrium, as confirmed by the prevalence of the forbidden component in the \ion{O}{vii} triplet. Two gas phases with different ionisation parameters are needed to reproduce the spectrum with a self-consistent photoionisation model, any contribution from a gas in collisional equilibrium being no more than 10\% of the total flux in the 0.35-1.55 keV band. When this self-consistent model is applied to the 0.5-10 keV band of the EPIC pn spectrum, a third photoionised phase is needed to account for emission lines with higher ionisation potential, although K$\alpha$ emission from \ion{S}{XV} and \ion{Fe}{xxvi} remains under-predicted.}
{}


\maketitle 

\section{ Introduction}

The X-ray spectrum of Compton-thick Seyfert 2 galaxies is dominated by reflection components, from both cold and ionised circumnuclear matter \citep[e.g.][]{matt00b}. The complete obscuration of the nuclear radiation, at least up to 10 keV, permits a clear view of these components that are heavily diluted in unobscured sources, often down to invisibility. Most of our knowledge of the circumnuclear matter, at least as far as their X-ray properties are concerned, is based on the brightest Compton-thick sources, like Circinus \citep[e.g.][]{matt99,Sambruna01b,mbm03}, NGC~1068 \citep[e.g.][]{kin02,matt04} and Mrk~3 \citep[e.g.][]{sako00b,bianchi05b,pp05}.

NGC 424 (a.k.a. Tololo 0109-383, z=0.0117) is one of the brightest Compton-thick Seyfert galaxies \citep[e.g.][]{matt00c}. Broad  H$\alpha$ and H$\beta$ lines were observed in polarised light \citep[e.g.][]{moran00}, where the polarisation degree is about 4\% after correction for starligh. \ion{Fe}{ii} emission and an extended 
(about 1 kpc) high ionisation nuclear emission line region (HINER), 70\% of which was however unresolved ($<$200 pc) have been discovered \citep{muray98}. HST/WFPC2 data showed the presence of a dust lane across the central part of the galaxy \citep{malk98}, which may help explain the observed $A_V\sim$1.4 to the NLR \citep{muray98}. The IRAS  colours are quite warm, suggesting that the IR emission is dominated by dust reprocessing of the nuclear radiation \citep[e.g.][]{matt00c}.

In X-rays, the ASCA spectrum shows a prominent iron line, a flat spectrum, and a high [\ion{O}{iii}]/F(2-10 keV) ratio, suggesting that the nucleus of NGC 424 should be absorbed by Compton--thick matter \citep{colbra00}. This result has been fully confirmed by the BeppoSAX observation \citep{matt00c,iwa01}, which measured an absorbing column of about 2$\times10^{24}$ cm$^{-2}$, for an estimated nuclear 2--10 keV luminosity of about 10$^{43}$ erg s$^{-1}$. NGC 424 was then observed by \textit{Chandra} and XMM-\textit{Newton} \citep{matt03}. Both observations were rather short (less than 10 ks each), but good enough to derive a few interesting properties of the source. The nuclear radiation was found to be reflected by both cold material (probably the inner wall of the torus) giving rise to the Compton reflection component and the iron K$\alpha$ fluorescent line, and by ionised matter, responsible for the soft X-ray emission. All these components are seen through a dust lane, responsible for the Balmer decrement and the absorption of the soft X-ray emission. Extended emission features in the Chandra observation are discussed in \citet{matt03}. We present here a new, long ($\sim 100 $ ks) XMM-\textit{Newton} observation of NGC 424, to investigate the properties of its circumnuclear regions in greater detail.

\section{Observations and data reduction}

The XMM-\textit{Newton} observation analysed in this paper was performed on 2008 December 7 (obsid 0550950101), with the EPIC CCD cameras, the pn \citep{struder01} and two MOS \citep{turner01}, operated in large window and medium filter, and the RGS cameras. Source extraction radii and screening for intervals of flaring particle background were performed with SAS 10.0.0 \citep{gabr04} via an iterative process which leads to a maximization of the Signal-to-Noise Ratio (SNR), similarly to that described in \citet{pico04}. After this process, the net exposure time was of about 104 ks for the pn, adopting an extraction radius of 31 arcsec and patterns 0 to 4. The background spectrum was extracted from source-free circular regions with a radius of 50 arcsec. Spectra were binned in order to over-sample the instrumental resolution by at least a factor of 3 and to have no less than 30 counts in each background-subtracted spectral channel. This allows the applicability of the $\chi^2$ statistics.

The RGS spectra were reduced following the guidelines in \citet{gb07}. Data reduction pipeline \texttt{rgsproc} was used, coupled with the latest calibration files available. We choose a fixed celestial reference point for the attitude solution, coincident with the NED optical nucleus of NGC 424. Source spectra were extracted in regions of the dispersion versus cross-dispersion and Pulse Invariant versus cross-dispersion planes, corresponding to 95$\%$ of the Point Spread Function (PSF) in the cross-dispersion direction. Background spectra have been generated using a sub-set of blank field observations, whose background counts matches the level measured during each individual RGS observation. The final net exposure times are about 123 ks for RGS1 and RGS2.

\section{Data analysis}

The adopted cosmological parameters are $H_0=70$ km s$^{-1}$ Mpc$^{-1}$, $\Omega_\Lambda=0.73$ and $\Omega_m=0.27$ \citep[i.e. the default ones in \textsc{xspec 12.5.1}:][]{xspec}. Errors correspond to the 90\% confidence level for one interesting parameter ($\Delta\chi^2=2.7$), if not otherwise stated. The RGS spectra were not re-binned and were analysed using the Cash-statistics \citep[][]{cash76}.

\subsection{RGS: phenomenological spectral analysis}

The soft X-ray spectrum of NGC 424 appears dominated by line emission, as commonly found in this class of sources \citep[e.g.][]{gb07}. As a first step, we performed phenomenological fits on $\simeq100$-bin spectral segments, using Gaussian profiles at the redshift of the source (z=0.0117), and a power law, both absorbed by the Galactic column density along the line of sight \citep[$N_H=1.8\times 10^{20}\ cm^{-2}$:][]{dl90}. Since the model used to fit the continuum is not very sensitive to the photon index $\Gamma$, due to the very limited band width of each segment, it has been fixed to $1$. Emission lines from H-like and He-like C, N, O, and Ne, as well as from the Fe L-shell, are all clearly detected (see Table~\ref{rgslines}). The spectrum also presents radiative recombination continua (RRC) from {O\,\textsc{vii}}, {O\,\textsc{viii}} and {C\,\textsc{vi}}. These features were fitted with the \texttt{REDGE} model in XSPEC. The width of the {O\,\textsc{vii}} RRC and {C\,\textsc{vi}} RRC are slightly resolved, allowing us to infer an electron temperature of $6^{+3}_{-3}$ eV for the former and $10^{+6}_{-4}$ eV for the latter. The same value inferred from the {O\,\textsc{vii}} RRC was adopted for the {O\,\textsc{viii}} RRC, because it could not be significantly constrained in the fit. All the observed transition energies are consistent with the theoretical values, but there is an hint of a systemic blueshift in some of the strongest emission lines (e.g. \ion{C}{vi} K$\alpha$ and the forbidden component of the \ion{O}{vii} triplet). On the other hand, no emission line is resolved, the tightest upper limits being those of the forbidden components of \ion{N}{vi} and \ion{O}{vii} K$\alpha$ ($\sigma<220$ and $270$ km s$^{-1}$).

\begin{table*}
\caption{\label{rgslines}Detected emission lines in the XMM-\textit{Newton} RGS spectra of NGC 424. Energies are in keV units, wavelengths in $\r{A}$, fluxes in $10^{-5}$ ph cm${}^{-2}$ s${}^{-1}$, kT in eV. Theoretical energies and wavelengths are from CHIANTI \citep{dere97,dere09}. The labelling for {Fe\,\textsc{xvii}} lines follows that of \citet{brown98}.}
\begin{center}
\begin{tabular}{lcccccccc}
& & & & & & & & \\
\hline
\hline
{\bfseries Line Id.} &   {\bfseries $\lambda_T$  } &{\bfseries $E_T$  }  &{\bfseries Energy  } & {\bfseries kT} & {\bfseries Fluxes } & & &  \\
& &  & & & (a)&(b) &(c) &(d)\\ 
\hline
 & & & & & & & &\\
{C\,\textsc{vi}}\ K$\alpha$& 33.736& 0.367  & $0.367^{+0.002}_{-0.001}$& - &$0.9^{+0.7}_{-0.9}$ & 1.10 & 0.5 & 0.6 \\
 & &  & & & & & &\\
 & 29.534&0.420 (f) &   $0.4196^{+0.0006}_{-0.0006}$&- &$2.4^{+0.7}_{-0.6}$ & 0.9 & 0.04 & 0.86 \\
{N\,\textsc{vi}}\ K$\alpha$ & 29.082&0.426 (i) & 0.426 &- &  $<0.8$& 0.2 & 0.03& 0.17 \\
 &28.787&0.431 (r) &  $0.4305^{+0.0007}_{-0.0003}$ & - &$<1.2$ & 0.3 &0.05 &0.25\\
 & & & & & & & &\\
{C\,\textsc{vi} RRC} &25.303 &0.490  & $0.490^{+0.001}_{-0.001}$ & $10^{+5}_{-3}$&$2.8^{+0.7}_{-0.7}$ &0.8&0.35&0.45\\
 & & & & & & & &\\
{N\,\textsc{vii}}\ K$\alpha$ & 24.781&0.500  & $0.500^{+0.005}_{-0.007}$ & -&$0.6^{+0.3}_{-0.3}$ & 0.3& 0.21& 0.09\\
 & & & & & & & &\\
  & 22.101&0.561 (f) &  $0.5618^{+0.0002}_{-0.0002}$ & -&$5.0^{+0.8}_{-0.8}$ &3.5 & 0.4&3.1\\
{O\,\textsc{vii}}\ K$\alpha$ & 21.807&0.569 (i)&0.569  & -&$<0.8$ & 0.9&0.1 &0.8\\
  & 21.602&0.574 (r)&$0.5735^{+0.0008}_{-0.0005}$    &-&$1.0^{+0.5}_{-0.5}$ & 0.81& 0.15&0.66\\
 & & & & & & & &  \\
{O\,\textsc{viii}}\ K$\alpha$ & 18.968&0.654  &$0.6535^{+0.0006}_{-0.0006}$ & -&$1.1^{+0.3}_{-0.3}$& 1.06&0.93&0.13 \\
 & &  & & & & & &\\
{O\,\textsc{vii}}\ K$\beta$ & 18.627&0.666 &$0.669^{+0.001}_{-0.001}$ & -&$0.5^{+0.2}_{-0.2}$& 0.04 & 0.02&0.02\\
 & &  & & & & & &\\
 {Fe\,\textsc{xvii}}\ M2 & 17.097&0.725 & \multirow{2}{*}{$0.7263^{+0.0002}_{-0.0005}$}& -& \multirow{2}{*}{$0.3^{+0.2}_{-0.2}$}& 0.003 &0.003& -\\
 {Fe\,\textsc{xvii}}\ 3G & 17.050&0.727 & & & & 0.02 &0.02&-\\
 & & & & & & & & \\
 {O\,\textsc{vii}}\ RRC & 16.769&0.739  &\multirow{2}{*}{$0.736^{+0.003}_{-0.003}$} & \multirow{2}{*}{$6^{+3}_{-3}$}& \multirow{2}{*}{$0.8^{+0.4}_{-0.3}$}& 0.6 &0.14&0.47\\
{Fe\,\textsc{xvii}}\ 3F& 16.777&0.739 & & & & 0.01 & 0.01&-\\
& & & &  & & & &\\
 {O\,\textsc{viii}}\ K$\beta$ & 16.005&0.775 & $0.773^{+0.002}_{-0.002}$ & -&$<0.3$ &0.04 &0.03 &0.01\\
 & &  & & & & & &\\
 {O\,\textsc{viii}}\ RRC & 14.228&0.871 &$0.871^{+0.004}_{-0.003}$ &$6^{+3}_{-3}$* &$0.6^{+0.2}_{-0.2}$ &0.59&0.53&0.07\\
 & &  & & & & & &\\
  &13.698& 0.905 (f)  &$0.904^{+0.001}_{-0.003}$ & -&$0.4^{+0.3}_{-0.2}$ & 0.40&0.22&0.18\\
  {Ne\,\textsc{ix}}\ K$\alpha$ & 13.551&0.915 (i) & 0.916 & -&$0.3^{+0.3}_{-0.2}$&0.13&0.07&0.06\\
   & 13.447 &0.922 (r) &$0.921^{+0.001}_{-0.001}$ & -&$0.5^{+0.3}_{-0.2}$& 0.17& 0.1&0.07 \\
 & & & & & & & &\\
 {Ne\,\textsc{x}}\ K$\alpha$ &12.133& 1.022 & \multirow{2}{*}{$1.022^{+0.002}_{-0.002}$} & \multirow{2}{*}{-} & \multirow{2}{*}{$0.7^{+0.2}_{-0.2}$} & 0.13 &0.13 &-\\
{Fe\,\textsc{xvii}} 4C & 12.119&1.023 &  &  & & & &\\
 & &  &  & & & & &\\
{Fe\,\textsc{xviii}} L & 10.524 &1.178 & \multirow{2}{*}{$1.177^{+0.005}_{-0.004}$} & \multirow{2}{*}{-} & \multirow{2}{*}{$0.4^{+0.3}_{-0.2}$}  & \multirow{2}{*}{0.006}& \multirow{2}{*}{0.006} &\multirow{2}{*}{-}\\
{Fe\,\textsc{xvii}} 7D & 10.500 &1.181 &  &  & &  &  & \\
& &  & & & & & &\\
{Mg\,\textsc{xi} (f)} & 9.168&1.352 & $1.370^{+0.003}_{-0.02}$ & - & $0.4^{+0.3}_{-0.1}$ & 0.05& 0.04& 0.01\\
& &  & & & & & &\\
\hline
\hline
\end{tabular}\\
\end{center}
(a) Calculated fluxes with the phenomenological analysis; (b) Line fluxes extrapolated from the RGS1+RGS2 best fit with CLOUDY; (c) Line fluxes arising from the component with higher photoionisation parameter ($\log U= 1.41^{+0.08}_{-0.07}$, $\log N_{H}=22.11^{+0.08}_{-0.11}$ ); (d) Line fluxes arising from the second photoionised phase ($\log U= 0.23^{+0.06}_{-0.03} $, $\log N_{H}=21.77 ^{+0.09}_{-0.07}$). 
\end{table*}

\begin{figure}
\includegraphics[width=9.cm,clip=]{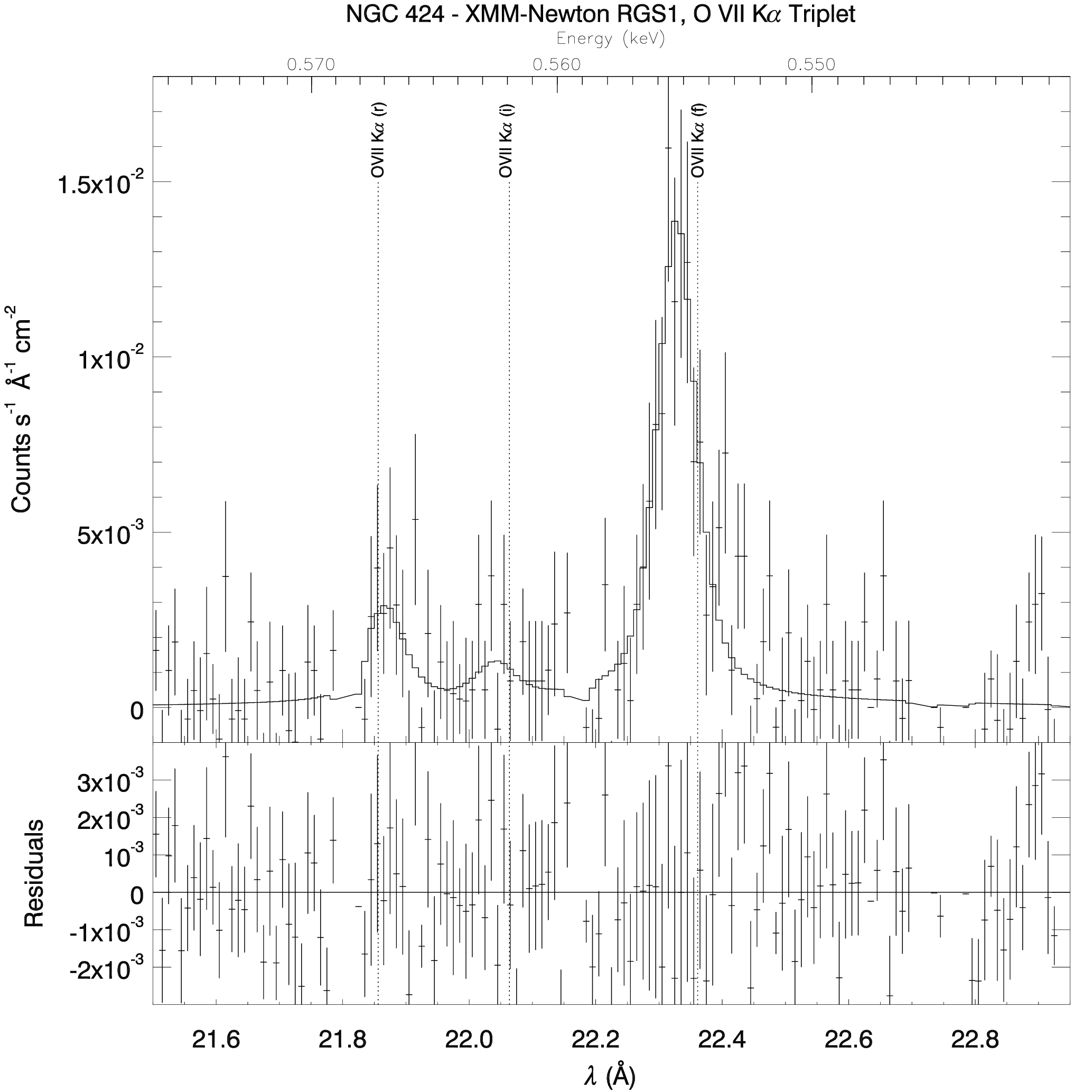}
\caption{\label{oviifit}RGS1 spectrum and best fit model for the phenomenological analysis: the \ion{O}{vii} triplet. The three transitions are labelled.}
\end{figure}

The analysis of the {O\,\textsc{vii}} triplet may provide precious information to determine whether the plasma is in photoionisation or collisional equilibrium. The three lines of the triplet are transitions between the n=2 shell and the n=1 ground-state shell, i.e.: the {\itshape resonance} line \textit{r}, corresponding to a ${}^1S_0- {}^1P_1$ transition, the {\itshape intercombination} line \textit{i} (actually a doublet: ${}^1S_0 - {}^3P_{2,1}$) and the {\itshape forbidden} line \textit{f} (${}^1S_0-{}^3S_1$). The ratio:
\begin{eqnarray}
G=\frac{f+i}{r}
\end{eqnarray}
is a good indicator of the predominant ionisation process. A weak resonance line compared to the forbidden or the intercombination lines is typical of plasmas dominated by photoionisation ($G>4$). On the contrary, a strong resonance line is produced in collision-dominated plasmas, where $G\sim1$ \citep[e.g.][]{pd00}. In the case of NGC 424, we can clearly infer that the gas is in photoionisation equilibrium, since $G\gtrsim 4.3$ (see Table~\ref{rgslines} and Fig.~ \ref{oviifit}).

The presence of sharp {O\,\textsc{vii}}, {O\,\textsc{viii}} and {C\,\textsc{vi}} RRC also provide evidence in favour of photoionisation equilibrium. Radiative recombination is the capture of a free electron, together with the emission of a photon with energy:
\begin{eqnarray}
\hbar \omega_n=E+\chi_n
\end{eqnarray}
where $E$ is the initial energy of the electron, and $\chi_n$ is the ionisation potential of the level into which the electron is captured. In the case of collisional ionisation equilibrium the ionisation potential $\chi_n$ is comparable to the initial energy of the electron $E\simeq kT$. As a result, the recombination radiation is broad, with a width of $\sim kT$. In the case of photoionisation, typically $kT\ll \chi_n$, and the recombination radiation is narrow. When observing this effect with the RGS on board XMM-Newton, the RRC in case of hot gas in collisional equilibrium is too broad to be visible. The presence of a narrow RRC is, therefore, a strong indicator of photoionisation. 

The {O\,\textsc{vii}} RRC (0.7393 keV) may be contaminated by the 3 F component of the {Fe\,\textsc{xvii}} L emission line at 0.7390 keV, since the instrument resolution does not allow us to discriminate them. Indeed, we detect an emission line at $\sim 0.7263$ keV, which can be securely identified as a blend of the 3G and M2 components of the same species, Fe \textsc{xvii}. As already discussed in \citet{bianchi10}, simulations with the \texttt{APEC} model in XSPEC showed that the ratio:
\begin{eqnarray}
\frac{Fe\textsc{ xvii 3F}}{Fe\textsc{ xvii (3G+M2)}}\simeq 0.44
\end{eqnarray}
for a wide range of temperatures. In our case the observed ratio between the line detected around 0.739 keV and that at 0.7263 keV is much higher, being $2.6^{+1.8}_{-1.9}$. We can therefore infer that most of the flux observed at 0.739 keV is due to the {O\,\textsc{vii}} RRC component.

\subsection{\label{rgscloudy}RGS: CLOUDY self-consistent model}

The high quality of the RGS spectrum, coupled with the results from the phenomenological spectral analysis, encouraged us to build a self-consistent model able to reproduce the whole spectrum, in a wavelength range from $8\ \r{A}$ up to $35\ \r{A} $. We produced a grid model for \textsc{xspec} using \textsc{cloudy} 08.00 \citep[last described by][]{cloudy}. It is an extension of the same model used in \citet{bianchi10}. The main ingredients are: plane parallel geometry, with the flux of photons striking the illuminated face of the cloud given in terms of ionisation parameter $U$ \citep{of06}; incident continuum modelled as in \citet{korista97}\footnote{We also used the \textit{IUE} UV fluxes reported by \citet{dunn06} to estimate an $\alpha_{ox}\simeq1.2$. Since this value is strongly dependent on the intrinsic X-ray flux of NGC~424, which is not directly observed, we decided to adopt a well-known standard continuum model, with an $\alpha_{ox}=1.4$, which is not significantly different from the one suggested by the data for this source.}; constant electron density $\mathrm{n_e}=10^5$ cm$^{-3}$; elemental abundances as in Table 9 of \textsc{cloudy} documentation\footnote{Hazy 1 version 08, p. 67: \url{http://viewvc.nublado.org/index.cgi/tags/release/c08.00/docs/hazy1_08.eps?revision=2342&root=cloudy}}; grid parameters are $\log U=[-2.00:4.00]$, step 0.25, and $\log N_\mathrm{H}=[19.0:23.5]$, step 0.1. Only the reflected spectrum, arising from the illuminated face of the cloud, has been taken into account in our model. We also produced tables with different densities ($\mathrm{n_e}=10^3-10^4$ cm$^{-3}$): all the fits presented in this paper resulted insensitive to this parameter, as expected since we are always treating density regimes where line ratios of He-like triplets are insensitive to density \citep{pd00}.

At first, we tried to fit the $8\ \r{A}-35\ \r{A} $ spectrum using a single photoionised phase. The obtained fit is rather good, with most of the lines detected, and with a Cash/dof$=4885/4500$. The best fit ionisation parameter is  $\log U=0.81^{+0.07}_{-0.15}$, with a column density $\log N_H=21.8^{+0.2}_{-0.3}$. Even if the fit was acceptable, we added another photoionised phase, to improve the residuals, for a final Cash/dof$=4865/4497$. The photoionisation parameters for this best fit are $\log U_1= 1.41^{+0.08}_{-0.07}$ and $\log U_2= 0.23^{+0.06}_{-0.03} $, while the column densities values are $\log N_{H_1}=22.11^{+0.08}_{-0.11}$ and $\log N_{H_2}=21.77 ^{+0.09}_{-0.07}$.  These values agree with similar studies performed on other Seyfert 2 objects \citep[e.g.][]{kin02,schurch04,bianchi10}. The total flux in the $0.35-1.55$ keV band is $(2.0^{+0.2}_{-0.8})\times 10^{-13}$erg cm${}^{-2}$ s${}^{-1}$, almost equally distributed between the two photoionised phases. A systemic blueshift of $-230^{+90}_{-120}\ $ km s$^{-1}$ is required by the fit, confirming the results found in the previous section. No significant improvement is obtained if different systemic velocities are considered for each of the photoionised phases.
Similar velocity blueshifts have been measured in the RGS spectrum of NGC 1068, as discussed in \citet{kin02}.\\
We tried to convolve our best fit model with a Gaussian smoothing (model \textsc{gsmooth} in \textsc{xspec}), to check if the emission lines are affected by a systemic broadening, and we get $\sigma=350^{+100}_{-150}\ $ km s$^{-1}$,  for a Cash/dof$=4858/4496$. This result agrees with the RGS phenomenological analysis, even if no emission line is individually resolved (see previous section). No significant improvement is obtained if we assume a different broadening for each of the photoionised phases.

The phenomenological analysis agrees with the self-consistent model fits, and the inferred fluxes of all the emission lines reproduced by the CLOUDY model are shown in Table~\ref{rgslines}. The two different components of the self-consistent model give different contributions to the whole spectrum. We can see in Table~\ref{rgslines} how the main contribution to the {O\,\textsc{vii}} K$\alpha$ line fluxes arises from the component with lower photoionisation parameter. However, in the case of {Ne\,\textsc{x}} K$\alpha$, the predicted flux falls short the observed one. We tried to introduce a further photoionised phase with a higher photoionisation parameter, but the fit is not sensitive to the new component and residuals due to {Ne\,\textsc{x}} K$\alpha$ line are still present. At almost the same energy, an emission line from {Fe\,\textsc{xvii}} is expected to be strong in collisionally ionised plasma. Since other {Fe\,\textsc{xvii}} lines are indeed detected in the spectrum, and their fluxes are all underestimated by our model, we tried to introduce a collisional component. The Cash$/$dof is $=4854/4494$ and the resulting kT is $0.54^{+0.45}_{-0.26}$ keV. The {Fe\,\textsc{xvii}} lines are marginally fitted, and the 0.35-1.55 keV flux due to the collisional phase is less than the 10$\%$ of the total flux. We conclude that there is no strong evidence for a collisional component in our spectrum. 

It is also interesting to note the underprediction by the model of the {O\,\textsc{vii}} K$\beta$ line and of the components of the {N\,\textsc{vi}} and {O\,\textsc{vii}} K$\alpha$ triplets, suggesting that the resonant lines are saturated. These inconsistencies could be due to the absence of turbulence velocities in our model. It was shown that the optical depths to resonant absorption can be more than 10 times lower when the gas is affected by strong turbulence \citep[greater than 300 km s$^{-1}$:][]{nfm99}. This effect would prevent the resonant lines to saturate up to higher column densities than in a gas not affected by turbulence, as assumed in our simple model.

\begin{figure*}
\includegraphics[width=18cm,clip=]{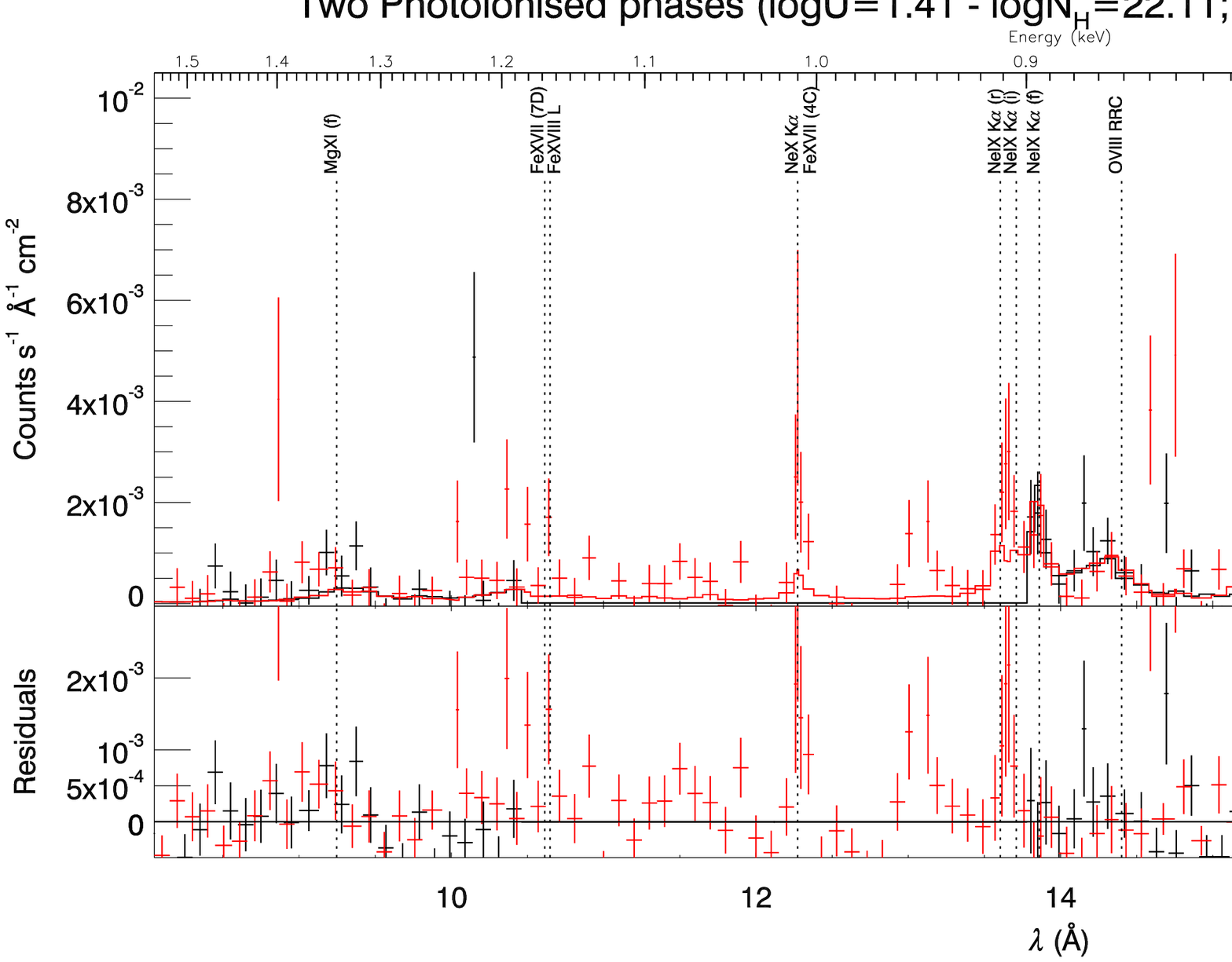}
\includegraphics[width=18cm,clip=]{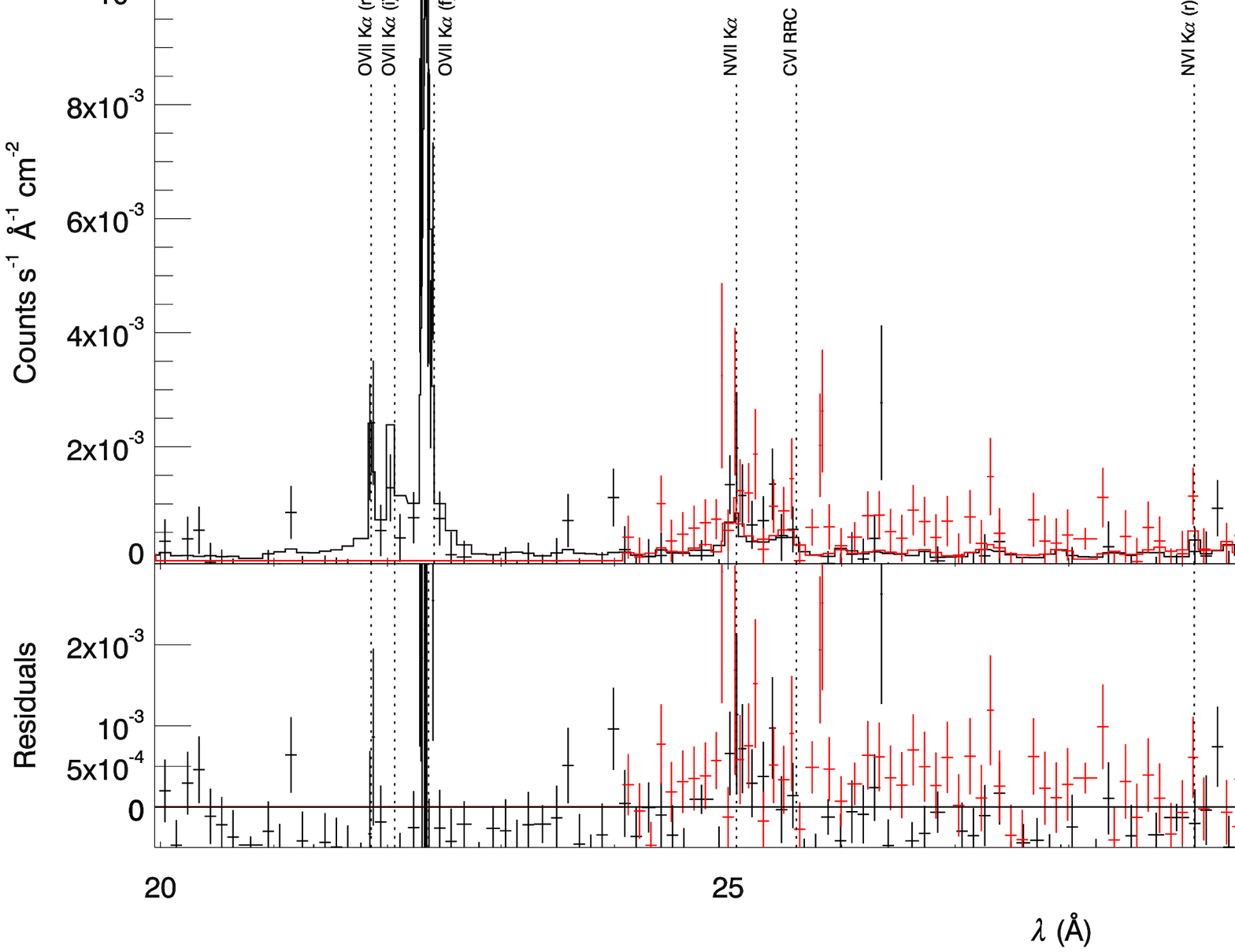}
\caption{\label{rgsplot}RGS1 (black) and RGS2 (red) self-consistent best fit (see text for detail). The 8-20 $\r{A}$ spectrum is shown in top panel, the 20-35 $\r{A}$ one in bottom panel.}
\end{figure*}

Apart for these lines, the residuals of the model with two photoionised phases are rather good, with the exception represented by some positive residuals around 13 and 33 $\r{A}$ (see Fig.~\ref{rgsplot}). Since the latter are significant only in the RGS1 spectrum we performed a separate analysis between the RGS spectra. It leads to a line detection only in the RGS1 spectrum ($3.8^{+2.0}_{-1.5}\times10^{-5}$ ph cm$^{-2}$ s$^{-1}$ at $0.3803^{+0.0004}_{-0.0001}$ keV), while in the RGS2 only an upper limit is obtained when the line energy is fixed at 0.3803 keV ($<0.6 \times10^{-5}$ ph cm$^{-2}$ s$^{-1}$). The same approach has been used to treat the residuals around 13 $\r{A}$, we found in the RGS2 spectrum a line at $0.947^{+0.004}_{-0.003}$ keV with a relative $0.3^{+0.2}_{-0.1}\times10^{-5}$ ph cm$^{-2}$ s$^{-1}$ flux. The latter result is fully consistent with a defective pixel in the RGS2 camera while since the identification of the former emission line would also be problematic, we conclude that the two previous detections are insecure\footnote{We note here that a line wavelength does correspond to a known defective pixel in one of the RGS cameras \url{(http://xmm.esac.esa.int/external/xmm_user_support/documentation/uhb/node59.html#3177)}}.

\subsection{EPIC pn spectral analysis}

\subsubsection{\label{510fit}5-10 keV phenomenological spectral analysis}

Prominent emission lines can be clearly observed in the hard part of the spectrum (5-10 keV). Following \citet{iwa01} and \citet{matt03}, it has been fitted with a model composed of a strongly absorbed ($\simeq10^{24}$ cm$^{-2}$) power-law with $\Gamma=2$, a pure cold reflection component (model \texttt{pexrav} in \textsc{xspec}, with the cosine of the inclination angle fixed to 0.45), and as many emission lines as required, all of them described by Gaussian profiles. The strongest detected emission line is the neutral Fe K$\alpha$ at 6.400 keV\footnote{Many of the following emission lines are indeed doublets, with a fixed intensity ratio between the two components of 1:2, and the given energies are the weighted mean \citep[based on the values reported by][]{{bea67}}. As already shown by \citet{yaq01} and \citet{bianchi05b}, given the small energy distance between the two components, the width of the line is unaffected by the modelling with a single Gaussian with current X-ray instruments.}, accompanied by a much fainter Fe K$\beta$ at 7.058 keV. K$\alpha$ emission from neutral Ni at 7.472 keV and \ion{Fe}{xxvi} at 6.966 keV are also detected, while only upper limits are found for the main components of the \ion{Fe}{xxv} K$\alpha$ triplet (see Table~\ref{highepic}). 

While all other lines are unresolved, the neutral Fe K$\alpha$ line width is $\sigma=81\pm12$ eV. Before discussing the physical meaning of this width, we investigated any possible instrumental effects. We analysed the calibration observations performed before and after our observation (obsid 0412990601 and 0510780201), and modelled the data with a power-law ($\Gamma=1$, any other choice does not affect the results) and three Gaussian lines, for the Mn K$\alpha$ doublet \citep[the separation of the lines fixed to be 0.0111 keV, and their flux ratio 1:2,][]{bea67} and the K$\beta$. All line widths were kept fixed to 0. The whole model was convolved with a Gaussian smoothing (model \textsc{gsmooth} in \textsc{xspec}), with free sigma. The best fit values for the \textsc{gsmooth} $\sigma$ were $62\pm6$ and $48\pm6$ eV, for the two data-sets. We therefore adopted \textsc{gsmooth} and the mean $\sigma=55$ eV to correct the pn spectrum of NGC 424.

\begin{figure*}
\includegraphics[width=9.cm,clip=]{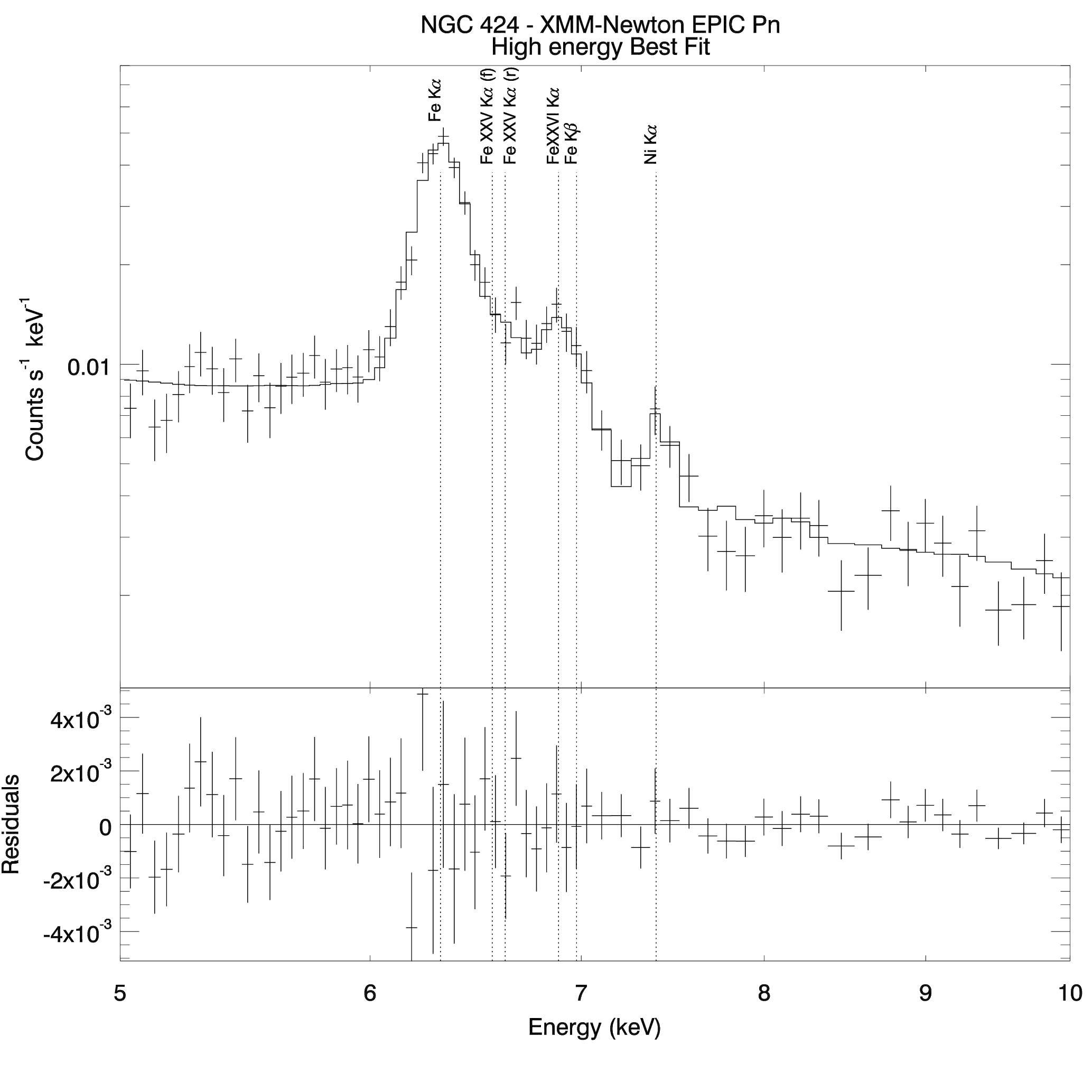}
\includegraphics[width=9.cm,clip=]{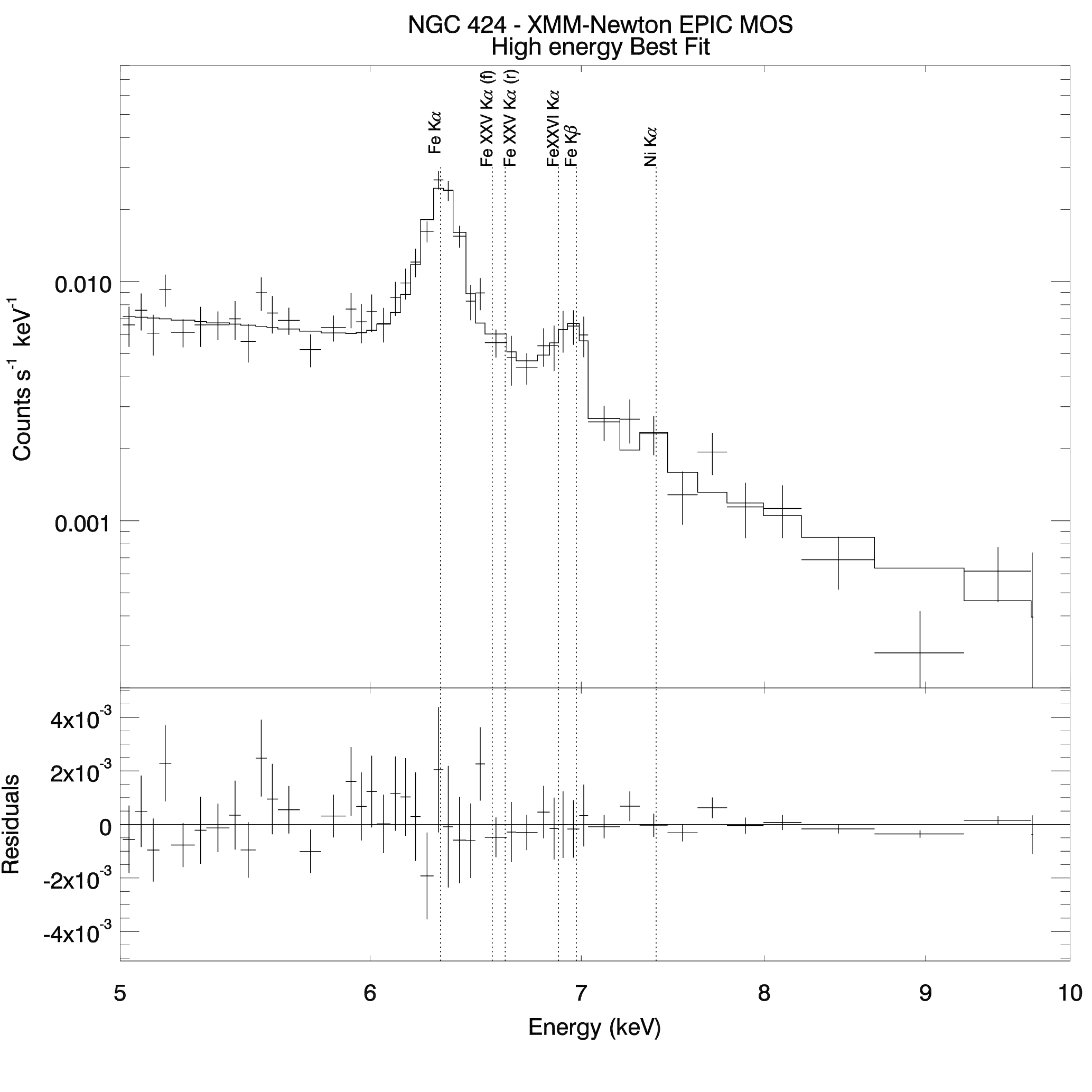}
\caption{\label{besthigh}Best-Fit of the hard-X (5-10 keV) EPIC pn (\textit{left}) and MOS (\textit{right}) spectrum.}
\end{figure*}

With this correction, the line width of the neutral Fe K$\alpha$ line is still resolved ($\sigma=48^{+16}_{-20}$ eV). We then added a Compton Shoulder (CS) redwards of the line core, as expected on theoretical ground, modelled as another Gaussian line, with energy fixed at 6.3 keV and $\sigma=40$ eV \citep[e.g.][]{matt02b}. Although the CS is not required by the data ($\Delta\chi^2=1$), its flux is $25\pm9\%$ of the flux of the narrow core, consistent with the expectations, and the narrow core of the Fe K$\alpha$ line is now unresolved ($\sigma<60$ eV). In Table~\ref{highepic} all the best fit results of the hard-X spectrum can be found and in Fig.~\ref{besthigh} the best fit plot is shown, with a resulting $\chi^2/dof=49/58$.

We checked the previous result by analysing the MOS spectra in the high energy band (5-10 keV). Lines' fluxes and energy centroids are fully consistent with the values presented in Table 3. The width of the neutral Fe K$\alpha$ line is $\sigma<50$ eV and the narrow core of the line is unresolved, for a best fit of $\chi^2=35/36$ dof.

If we consider the neutral Fe K$\alpha$ line resolved, without any Compton Shoulder redwards of the line core, we get a $\sigma=48^{+16}_{-20}$ eV. With this result considerations on the geometry of the inner radius of the molecular torus can be drawn. If we assume that the torus has a Keplerian motion around the central super massive black hole, it is easy to show that the expected FWHM of a line produced in its inner walls should follow the relation $2v_k\sin i \simeq 1300 (M_8/r)^{1/2}\sin i$ km s${}^{-1}$, where the radius is expressed in parsec, the mass is in $10^8$ M$_{\odot}$ units and $i$ is the angle between the torus axis and the line of sight. The black hole mass of NGC 424 is estimated by means of stellar velocity dispersion to be $6.02\times 10^7$ M$_{\odot}$ \citep[e.g.][]{biangu07}. This means that the expected FWHM for the iron line in this source is 1010 $r^{-1/2}\sin i$ km s${}^{-1}$. If the FWHM=$5300{}^{+1800}_{-2200}$ km s${}^{-1}$ we measure with the EPIC pn is due to Doppler broadening\footnote{A proper modelling of the line profile should take into account that the rotating torus velocity distribution is double-peaked, but the quality of the data does not allow us to explore it in detail.}, we can estimate the inner radius of the torus: $r=0.04{}^{+0.06}_{-0.02} \sin^2 i$ pc. For different choices of $i$ of 30${}^{\circ}$ or 60${}^{\circ}$ we get central values of 0.01 and 0.03 pc, respectively, for the inner radius.
  
\begin{table}
\caption{\label{highepic}EPIC pn phenomenological fit (5-10 keV). See text for details.}
\begin{center}
\begin{tabular}{lccc}
\hline
\hline
& &\\
{\bfseries Id.} & \textbf{Energy} & \textbf{Flux} & \textbf{EW}\\
 \hline
& & &\\
Fe K$\alpha$ & $6.40^{+0.02}_{-0.01}$ & $10.8^{+1.8}_{-1.1}$ & $810^{+130}_{-80}$\\
& & &\\
Fe K$\beta$ & $7.058^{*}$ & $<1.0$ & $<90$\\
& & &\\
Fe XXVI K$\alpha$ & $6.966^{*}$ & $1.5^{+0.9}_{-1.0}$ & $120^{+70}_{-80}$\\
& & &\\
Ni K$\alpha$ & $7.472^{*}$ & $1.3\pm0.5$ & $170\pm70$\\
& & &\\
Fe XXV K$\alpha$ (f) & $6.637^{*}$ & $<1.9$ & $<140$\\
& & &\\
Fe XXV K$\alpha$ (r) & $6.700^{*}$ & $<2.1$ & $<150$\\
& & &\\
Compton Shoulder & $6.300^{*}$ &$2.7\pm1.0$ & $210\pm80$\\
& & &\\
\hline
\hline
\end{tabular}
\end{center}
$^*$ fixed at the theoretical value.

Energies are in keV, fluxes in $10^{-6}$ ph cm$^{-2}$ s$^{-1}$, and EWs in eV.
\end{table}

We can use the best fit results to provide estimates on the neutral gas parameters, such as the Fe and Ni abundance, and the ionisation state. The iron abundance is measured by the depth of the iron edge in the Compton reflection continuum, and it is measured with respect to the elements responsible for the photoabsorption below the edge, mainly oxygen and neon but with not negligible contributions from magnesium, silicon and sulphur. Leaving this parameter free to vary in the best fit, we can measure it with high statistical precision, i.e. $A_{Fe}=1.00{}^{+0.25}_{-0.17}$ in solar units \citep[e.g][]{ag89}. From the observed Ni K$\alpha$ to Fe K$\alpha$ line fluxes, we can instead estimate the relative abundances of the two elements. The expected ratio ranges from $0.03$ to $0.045$, depending on the inclination angle and the incident power-law index, as discussed in \citet{mbm03}. In our case, we measure $0.12\pm0.05$ (considering only the flux of the iron K$\alpha$ core line, excluding the CS), significantly higher than the expected one, indicating a nickel-to-iron overabundance by a factor $\simeq2$. This result agrees fully with previous studies on high energy spectra performed on similar Compton thick Seyfert 2 galaxies, such as NGC~1068 \citep{matt04} and Mrk~3 \citep{bianchi05b}. The presence of an Fe K$\beta$ line at 7.058 keV is not statistically required, leading to an FeK$\beta$/FeK$\alpha$ (core only) ratio  $\lesssim 0.10$. The expected value is higher, from 0.155 to 0.16 \citep{mbm03}, suggesting a moderate ionisation of the gas responsible for iron fluorescence.

The detection of the \ion{Fe}{xxvi} K$\alpha$ emission line requires the presence of another gas with very different physical properties from the one responsible for neutral fluorescence and the Compton reflection continuum. Given the limited statistics and the upper limits on the \ion{Fe}{xxv} K$\alpha$ triplet, both a pure photoionised Compton-thin material ($\log U=3.5^{+0.4}_{-0.3}$ and $\log N_H=20.0^{+0.4}_{-0.3}$, adopting the same model described in Sect.~\ref{rgscloudy}) and a plasma in collisional equilibrium (kT=$10.9^{+4.1}_{-3.6}$ keV, model \textsc{apec})  reproduce the emission line equally well, with a final $\chi^2$ statistically equivalent to the phenomenological best fit.

\subsubsection{Broad band EPIC pn spectrum (0.5-10 keV)}

\begin{figure}
\begin{center}
\epsfig{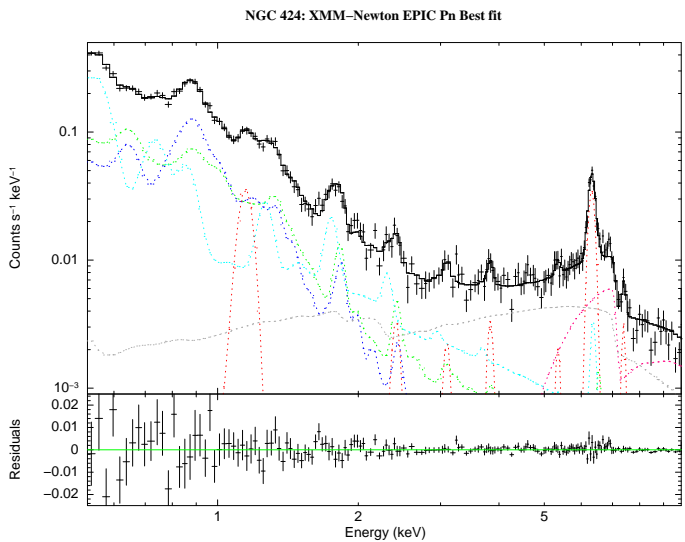}
\caption{\label{broadband}XMM-\textit{Newton} EPIC pn 0.5-10 keV best fit and residuals. All the model components are shown: pure Compton reflection (grey), absorbed primary powerlaw (magenta), low photoionised component (light blue), medium photoionised component (blue), high photoionised component (green), Gaussian lines (red). See text for details.}
\end{center}
\end{figure}

Taking into account the results from the soft X-ray RGS spectrum, we finally performed a self-consistent fit in the whole EPIC pn band (0.5-10 keV). The model for the high energy spectrum is the same that we adopted in the previous section, including a photoionisation component needed to reproduce for the \ion{Fe}{xxvi} K$\alpha$ emission line. As for the soft X-ray emission spectrum, we adopted the same model which successfully reproduce the RGS spectrum, consisting in two photoionisation components (we fixed for all the components the same velocity shift measured in the RGS analysis). The overall fit is marginally acceptable ($\chi^2=256/167$ dof), due to some residuals suggesting the presence of emission lines not reproduced by our model.

Indeed, the addition of a line at $1.16\pm0.01$ keV is significantly required ($\Delta\chi^2=51$), its flux of $3.4\pm0.8\times10^{-6}$ ph cm$^{-2}$ s$^{-1}$, agrees with the \ion{Fe}{xviii} and \ion{Fe}{xvii} L blend observed in the phenomenological analysis of the RGS spectrum (see Table~\ref{rgslines}). Another emission line, at $2.44\pm0.02$ keV, with flux $1.1\pm0.4\times10^{-6}$ ph cm$^{-2}$ s$^{-1}$, improves the fit ($\Delta\chi^2=5$). It can be identified with a blend of the \ion{S}{XV} triplet (2.430, 2.448, 2.461 keV)\footnote{We excluded that this line could be an artifact of inadequate EPIC calibration near the gold M-edge by applying a gain fit to the data.}. More marginal detections are found at $3.13\pm0.05$ keV ($\Delta\chi^2=4$), identified with the \ion{Ar}{xvii} K$\alpha$ triplet (3.104, 3.125, and 3.140 keV), $3.88\pm0.04$ keV ($\Delta\chi^2=9$), i.e. the \ion{Ca}{XIX} K$\alpha$ triplet (3.861, 3.883, and 3.888 keV), and at $5.37\pm0.07$ keV ($\Delta\chi^2=4$), consistent with K$\alpha$ fluorescence from neutral Cr (5.412 keV). The emission line of another high ionisation potential species is under-predicted, i.e. \ion{Fe}{xxvi}. This suggests that the gas component with the highest ionisation parameter is likely more complex, and cannot be easily modelled with a single phase, which is not able to reproduce all the observed emission.

The final fit is good ($\chi^2=183/162$ dof). The best fit values for the three photoionisation components are\footnote{Lower limits indicate that the parameters cannot be constrained above the maximum values considered in our grid model.}: $\log U_1=2.1^{+0.1}_{-0.1}$, $\log N_{H1}>23.3$; $\log U_2=1.6^{+0.1}_{-0.1}$, $\log N_{H2}=20.4\pm0.5$; $\log U_3=-1.21^{+0.06}_{-0.03}$, $\log N_{H3}=22.4^{+0.3}_{-0.2}$. The intermediate component agrees fully with the one found in the RGS analysis (see Sect.~\ref{rgscloudy}), both the least and the most ionised component have lower ionisation parameters than those extracted from the analysis in the bands where they dominate. This is due to the not negligible contribution of the high ionisation component to the soft X-ray emission, reflecting the possibility that, as already mentioned above, the modelling in three separate phases is too simple, and more complex geometrical and physical dependencies between these components are needed in order to reproduce at the same time all the observed features in the broad band.

As a self-consistent check, we extrapolated our X-ray best fit for the photoionised components to the optical band, in order to compare the predicted [\ion{O}{iii}] flux with the observed one. Our model produces an excess in this emission line, by a factor of $16\pm8$, taking into account the errors on our best fit. The comparison was made with the reddening-corrected [\ion{O}{iii}] flux \citep{muray98}, so the discrepancy is intrinsic, and not due to the reddening of the NLR. Moreover, no additional neutral absorption at the redshift of the source is required in our best fit ($N_H<1.6\times10^{20}$ cm$^{-2}$), at odds with the measured $A_V\sim$1.4 to the NLR \citep{muray98}, and differently from what found by \citet{matt03}, although with a much simpler modelling of the soft X-ray emission. The only contribution to the [\ion{O}{iii}] flux in our best fit model comes from the photoionised component with the lower ionisation parameter. In the X-ray band, this phase produces mostly \ion{O}{vii} emission, but also strong emission lines from quasi-neutral S, Si, and Mg (see also Fig.~\ref{broadband}). A significant contribution to these lines likely comes from the Compton-thick material producing the neutral iron K$\alpha$ line, but they are completely fitted by the low photoionised phase in our model. Consequently, its flux may be higher than expected, producing part of the observed discrepancy in the [\ion{O}{iii}] emission line. It should also be noted that, in order to perform a detailed comparison between the optical and the X-ray emission of the NLR, a much better modelling of the spectral energy distribution of NGC 424 should be adopted in the \textsc{cloudy} models, but this is beyond the scopes of this paper.

Finally, the intrinsic 2-10 keV luminosity of NGC 424 is $4.2\times10^{42}$ erg s$^{-1}$, absorbed by a neutral column density of $1.1\pm0.2\times10^{24}$ cm$^{-2}$, it roughly agrees with the value measured by \citet{iwa01}.  The 2-10 keV luminosity of the Compton reflection component is $9.6\times10^{40}$ erg s$^{-1}$, about an half of the reflection expected from a slab subtending at 2$\pi$. Note that the absorption includes Compton scattering only along the line-of-sight (model \textsc{cabs}), so the derived intrinsic luminosity can be considered an upper limit to the real one, depending on the covering factor of the absorber, and, thus, the reflection fraction is likely higher than $\simeq0.5$. Measuring the reflection fraction in Compton thick AGN is obviously impossible besides the few cases (such as NGC 424) in which the intrinsic continuum is not totally obscured below 10 keV. However R$\simeq0.5$ is in line with the typical reflection fraction of the distant reflection component observed in unobscured Seyfert 1 galaxies \citep[e.g.]{nan07} supporting the standard view that distant reflection in both obscured and unobscured AGN arises in one and the same medium with similar global covering factor.

Adopting the X-ray bolometric correction presented in \citet{mar04}, we can estimate the bolometric luminosity of NGC 424 from the intrinsic 2-10 keV as $L_{bol}=5.9\times10^{43}$ erg s$^{-1}$. This value agrees with the one that we can derive from the [\ion{O}{iii}] luminosity, once corrected for reddening, extracted from the optical data in \citet{muray98}, and adopting the bolometric correction proposed by \citet{lamastra09}: $L_{bol}=5.1\times10^{43}$ erg s$^{-1}$. Assuming the BH mass already mentioned in Sect.~\ref{510fit}, the accretion rate of the source is therefore $\dot{m}\simeq6.7-7.5\times 10^{-3}$ in Eddington units.

\section{Conclusions}

We presented the XMM-\textit{Newton} RGS and EPIC pn spectra of a long observation of the Seyfert 2 galaxy, NGC 424. The high-energy part of the spectrum confirmed its nature as a Compton-thick source, dominated by a pure neutral Compton reflection component, and a neutral iron K$\alpha$ line with a very large EW, together with strong K$\alpha$ emission from neutral Ni, suggesting a significant Ni/Fe overabundance. From the width of the neutral iron K$\alpha$ line we estimated the inner radius of the torus: $r=0.04{}^{+0.06}_{-0.02} \sin^2 i$ pc, where $i$ is the inclination angle between the torus axis and the line of sight. The presence of \ion{Fe}{xxvi} K$\alpha$ emission can be reproduced equally well by a plasma in collisional or photoionisation equilibrium, given that only upper limits can be recovered on emission from \ion{Fe}{xxv}.

The soft X-ray RGS spectrum of NGC 424 comes mostly from line emission from H-like and He-like C, N, O, and Ne, as well as from the Fe L-shell, as commonly found in obscured AGN \citep[e.g.][]{gb07}. The presence of narrow RRC from \ion{O}{viii}, \ion{O}{vii}, and \ion{C}{vi} (the latter two with resolved widths corresponding to temperatures around 5-10 eV) is a strong indication of a low-temperature gas, which must therefore be in photoionisation equilibrium to have such an high ionisation state. This is confirmed by the prevalence of the forbidden component in the \ion{O}{vii} triplet. Indeed, a self-consistent photoionisation model well reproduces the RGS spectrum, adopting two gas phases with different ionisation parameters. This gas is likely to be identified with the ionisation cones of the NLR, as suggested for most Seyfert 2 galaxies \citep[e.g.][]{kin02,schurch04,bianchi10}, although a contribution from inner regions (i.e. the torus, the BLR) cannot be ruled out. Any contribution from a gas in collisional equilibrium should not exceed 10\% of the total flux in the 0.35-1.55 keV band.

When this self-consistent model is applied to the 0.5-10 keV band of the EPIC pn spectrum, a third photoionised phase is needed to account for emission lines with higher ionisation potential, although K$\alpha$ emission from \ion{S}{XV} and \ion{Fe}{xxvi} remains under-predicted.

\acknowledgement
AM, SB, GM and EP acknowledge financial support from ASI (grant I/088/06/0). We would like to thank Matteo Guainazzi for very useful discussions on EPIC calibration issues.
\bibliographystyle{aa}
\bibliography{sbs} 

\end{document}